\documentclass{elsart}
\usepackage{graphicx}

\begin{document}

\begin{frontmatter}

\title{Global features of proton-neutron interactions and symmetry
 energy}
 
\author[higashi]{K. Kaneko} and 
\author[Sawara]{M. Hasegawa}

\address[higashi]{Department of Physics, Kyushu Sangyo University,
 Fukuoka 813-8503, Japan}
\address[Sawara]{Laboratory of Physics, Fukuoka Dental College,
 Fukuoka 814-0193, Japan}
 
\begin{abstract}
 We study global features of proton-neutron ({\it p-n}) interactions and
 symmetry energy over a wide range of nuclei, using a schematic model
 interaction with four forces proposed recently.
 Calculations are performed by the BCS approximation in $N,Z$=20-50
 and $N,Z$=50-82 regions.  The experimental double differences of
 binding energies and symmetry energy are reproduced quite well.
 It is shown that the isoscalar {\it p-n} interactions with all $J$ are
 indispensable for explaining the binding energies of not only
 $N\approx Z$ but also $N>Z$ nuclei in the $A=$40-160 region. 

\vspace*{3mm}
\leftline{PACS: 21.10.Dr;21.10.Hw;21.60.-n;21.60.Cs}
\end{abstract}

\end{frontmatter}

Corresponding author:

K. Kaneko\\
Department of Physics, Kyushu Sangyo University,
 Fukuoka 813-8503, Japan\\
e-mail address: kaneko@phys.kyusan-u.ac.jp\\
Fax: Japan 092-673-5899

\newpage

  In recent years, the study of proton-neutron ({\it p-n}) interactions
 has been a hot topic in the proton rich side of nuclide table.
 The {\it p-n} interactions are expected to become strong in $N \approx Z$
 nuclei where valence protons and neutrons in the same orbits have
 large overlaps of the wavefunctions, and to play very important roles
 in the structure of these nuclei \cite{Goodman}.
 The double differences of binding energies, which were investigated
 for a different purpose before \cite{Janecke1,Jensen,Dussel},
 have been recognized as useful measures to study the p-n interactions
 \cite{Zhang,Brenner,Kaneko3}.  Their observed values display
 characteristic behaviors near $N=Z$ and reveal properties of the p-n
 interactions.  We have analyzed what {\it p-n} interactions cause the
 characteristic behaviors in a previous paper \cite{Kaneko3}. 
 One of our conclusions is that the double differences of binding
 energies trace a curve $40/A$ on the average and the average value
 is attributed to the isoscalar {\it p-n} interaction, almost to the
 $J$ independent isoscalar {\it p-n} force.
 The investigation, however, is carried out in the $N \approx$ Z 
 nuclei where a large $j$ subshell dominates, i.e., $f_{7/2}$ and
 $g_{9/2}$ shell nuclei.  On the other hand, the curve $40/A$ observed
 in the double differences of binding energies continues till heavy
 nuclei.  This suggests that considerably strong isoscalar {\it p-n}
 interaction persists over a wide range of nuclei with $N>Z$.

  The competition of the isoscalar {\it p-n} interaction with the isovector
 one is observed in the energy spectrum as well as the binding energy
 \cite{Eng1,Lang,Eng2,Rud,Dean,Kaneko1,Vogel,Poves,Kaneko2,Hasegawa}.
 It was discussed that the degeneracy of $T=0$ and $T=1$ states in
 odd-odd $N=Z$ nuclei are explained by a near balance of the isovector
 pairing and symmetry energy \cite{Vogel}.  
 The experimental data indicate that the symmetry energy accompanied
 by the so-called Wigner energy is 
 proportional to $T(T+1)$ with a common
 coefficient $a(A)$, i.e.,  $E_{sym}+E_W=a(A)T(T+1)/A$
 where $T=|T_Z|=|N-Z|/2$.  It has been shown that the Wigner energy is
 originated in the isoscalar {\it p-n} interaction \cite{Satula1,Satula2}
 but cannot be solely explained in terms of the $J=1$ isoscalar {\it p-n}
 interaction \cite{Poves,Satula2}. These investigations have suggested
 the importance of isoscalar {\it p-n} interactions with various $J$ in 
 the symmetry energy and Wigner energy.  This is confirmed in the single
 $j$ shell model calculation for $f_{7/2}$ shell nuclei in our previous
 paper \cite{Kaneko3}.  Our model reproduces very well the empirical
 symmetry energy coefficient $a(A)$.  The result indicates that
 the symmetry energy coefficient is explained by the $J$ independent
 isoscalar {\it p-n} force and $J=0$ isovector pairing force.
 In this letter, we extend our discussion to a wide range of nuclei.
 The empirical formula of Seeger \cite{Seeger1,Seeger2} or Duflo and
 Zucker \cite{Duflo} for $a(A)$ has a mass dependence
 $(1-\eta A^{-1/3})$.
 This $A$ dependence combined with the curve 40/A of the average double
 differences of binding energies will give a good guide to determine
 the $A$ dependence of our isoscalar {\it p-n} force strength.
 
 Our previous study \cite{Kaneko3} has been carried out using
 a schematic interaction composed of four forces which is extended from
 the pairing plus $QQ$ force.  This schematic interaction reproduces well
 the low-lying states and binding energies of $N \approx Z$ nuclei
 where a large $j$ subshell dominates, and also various nuclear
 properties, such as the double differences of binding energies,
 symmetry energy, Wigner energy, odd-even mass difference and
 two-proton separation energy. The {\it p-n} interactions
  play leading roles
 in these properties, especially the $J$ independent isoscalar {\it p-n}
 force is essential for the properties which we consider here.
 The schematic interaction with four force parameters is easy to extend
 the model space.  In this letter, we investigate systematics of
 the double differences of binding energies and symmetry energy in both
 of $N \approx Z$ and $N>Z$ nuclei with $N,Z=$20-50 and $N,Z=$50-80.
 We perform this many $j$ shell calculation in the BCS approximation.
 
 The Hamiltonian we proposed in Ref. \cite{Hasegawa} has an
 isospin-invariant form as follows:
\begin{eqnarray}
 & & H = H_{\rm sp} + H_{{\rm P}_{0}} + H_{\rm QQ} + H_{{\rm P}_{2}}
  + H^{\tau =0}_{\pi \nu}, \label{eq:1} \\
 & {} & H_{\rm sp}= \sum_{\alpha \rho} (\epsilon_a - \lambda)
  c^\dagger_{\alpha \rho} c_{\alpha \rho}, \label{eq:2} \\
 & {} & H_{{\rm P}_{J}}= - {1 \over 2} g_{J} \sum_{M \kappa}
  \sum_{a \leq b} \sum_{c \leq d} p_J(ab) p_J(cd)
  A^\dagger_{JM1\kappa}(ab) A_{JM1\kappa}(cd),   \label{eq:3} \\
 & {} & H_{\rm QQ}=-{1 \over 2} \chi \sum_M \sum_{ab \rho}
  \sum_{cd \rho^\prime} q(ab) q(cd) : B^\dagger_{2M \rho}(ab)
  B_{2M \rho^\prime}(cd):  \label{eq:4} \\
 & {} & H^{\tau =0}_{\pi \nu}= - k^{0}\sum_{a \leq b} \sum_{JM}
    A^\dagger_{JM00}(ab) A_{JM00}(ab),
   \label{eq:5}
\end{eqnarray}
with
\begin{eqnarray}
 A^{\dagger}_{JM \tau \kappa}(ab) = { [ c^{\dagger}_{\alpha \rho} 
 c^{\dagger}_{\beta \rho^\prime} ]^{\tau \kappa}_{JM}  \over
  \sqrt{1 + \delta_{ab} } }, \quad
 B^\dagger_{JM \rho}(ab) = [ c^\dagger_{a \rho} c_{b \rho} ]_{JM}.
  \label{eq:6} 
\end{eqnarray}
 Here $p_{0}(ab)$=$\sqrt{ (2j_a+1) } \delta_{ab}$
 and $p_{2}(ab)$=$q(ab)/b^{2}$=$(a\|r^2Y_2/b^{2}\|b)/\sqrt{5}$ 
 where $b^2$ is the harmonic-oscillator range parameter. 
 This Hamiltonian has only four parameters, $g_{0},g_{2},\chi,$ and
 $k^{0}$.

 We consider the model space (1$f_{7/2}$, 2p$_{3/2}$,
 1$f_{5/2}$, 2$p_{1/2}$, 1$g_{9/2}$) for $N,Z=20-50$ region and 
 (2$d_{5/2}$, 1$g_{7/2}$, 3$s_{1/2}$, 2$d_{3/2}$, 1$h_{11/2}$) for
 $N,Z=50-82$ region. The lowest single-particle energy in each region,
 is determined as $\epsilon_{f_{7/2}}-\lambda=B(^{41}{\rm Ca})
 -B(^{40}{\rm Ca})=-8.3633$ MeV and $\epsilon_{d_{5/2}}-\lambda
 =B(^{103}{\rm Sn})-B(^{102}{\rm Sn})=-10.265$ MeV from the
 experimental binding energies. The other single particle energies
 are chosen so as to be the same level spacings as those used by
 Kisslinger and Sorensen \cite{Kisslinger}. The pairing force strength
 $g_{0}$ and the quarupole force strength $\chi$ are ajusted to fit
 approximately the experimental odd-even mass differences and the
 lowest $J=2^{+}$ energies in quasiparticle Tamm-Dancoff approximation,
 respectively, 
\begin{eqnarray}
 g_{0}=24A^{-1}, \qquad \chi=350A^{-5/3}b^{-4} \qquad \mbox{ for } 
 N,Z=20-50 \mbox{ region}, \label{eq:8} \\
 g_{0}=24A^{-1}, \qquad \chi=450A^{-5/3}b^{-4} \qquad \mbox{ for } 
 N,Z=50-82 \mbox{ region}. \label{eq:9} 
\end{eqnarray}
The quadrupole pairing force strength $g_2$ can be determined by
 the same relation $g_{2}=0.2g_{0}$ as that used by Hara {\it et al.}
 \cite{Hara}.  We shall fix the strength $k^0$ of the isoscalar 
 {\it p-n} force (\ref{eq:5}) in the following discussion.

 Let us first analyze the experimental values of the $m$th double
 difference of binding energies defined by the following equation
 \cite{Kaneko3},
\begin{eqnarray}
\delta V^{(m)}(Z,N)=\delta^{(m)}B(Z,N), \label{eq:10}
\end{eqnarray}
where $B(Z,N)$ is the nuclear binding energy and the operator
 $\delta^{(m)}$ is defined as
\begin{eqnarray}
\delta^{(m)}f(Z,N) &=&-\frac{1}{m^{2}}[f(Z,N)-f(Z,N-m)  \nonumber \\
   & {} & -f(Z-m,N)+f(Z-m,N-m)]. \label{eq:11}
\end{eqnarray}

 Figure 1(a) shows the plot of $\delta V^{(1)}(Z,N)$ as a function of
 $A=N+Z$ for nuclei in the mass region $A$=16-210. We can see two
 separate groups in the figure, namely, one is for the even $A$ nuclei
 (dots) and the other is for the odd $A$ nuclei (crosses).
 The patterns of dots and crosses show shell effect at $Z$ or
 $N=$28, 40, 50, 82 and are symmetric with respect to the average
 curve $40/A$.  The double difference of binding energies
 $\delta V^{(1)}(Z,N)$ can be approximately written as 
\begin{eqnarray}
\delta V^{(1)}(Z,N)= \frac{40}{A} + (-1)^{A}I', \label{eq:12}
\end{eqnarray}
where $I'$ denotes a deviation from the average $40/A$. 
 The large values of $\delta V^{(1)}(Z,N)$ for even-$A$ nuclei
 (circled dots) at $N=Z$ are striking below $A=80$.
 The data of another double difference of binding energies
 $\delta V^{(2)}(Z,N)$ are plotted in Fig. 1(b) as a function of
 $A=N+Z$. Our definition of $\delta V^{(2)}$ has the opposite sign
 to Brenner's one \cite{Brenner}. 
 The $\delta V^{(2)}$ values show a different behavior from the
 $\delta V^{(1)}$ ones. We see large scatters of dots and crosses for
 $A<80$, which show $\delta V^{(2)}$ of $N \approx Z$ nuclei.
 The values of $\delta V^{(2)}$ at $N=Z$ are especially large as
  that of $\delta V^{(1)}$. 
 If we neglect the scattered dots and crosses in $N \approx Z$ nuclei,
 $\delta V^{(2)}$ varies rather smoothly. This smooth trend is clear
 for $A>80$ and continues up to heavy nuclei. This is due to the fact
 that there are no stable $N \approx Z$ nuclei with $A>80$.
 The systematic decrease of $\delta V^{(2)}$ with increasing mass $A$
 can be traced by the same average curve $40/A$ as that of
 $\delta V^{(1)}$. The deviations from the curve $40/A$ are small and
 shell effect disappears for $\delta V^{(2)}$.
 In Ref. \cite{Kaneko3}, we analyzed $\delta V^{(1)}$ and
 $\delta V^{(2)}$ for probing the {\it p-n} interactions in $N\approx Z$
 nuclei of $g_{9/2}$ and $f_{7/2}$ subshells. The analysis reveals
 different contributions of isoscalar and isovector {\it p-n} interactions
 to the double differences of binding energies. The average curve
 $40/A$ of $\delta V^{(1)}$ and $\delta V^{(2)}$ is attributed almost
 to the isoscalar {\it p-n} force $H^{\tau =0}_{\pi \nu}$ introduced, while
 the {\it p-n} part of $P_{0}+QQ+P_{2}$ force produces the deviation from
 the curve $40/A$ in $\delta V^{(1)}(Z,N)$. 

 Since $\delta V^{(2)}$ comes from the $J$ independent isoscalar {\it p-n}
 force $H^{\tau =0}_{\pi \nu}$, the average curve $40/A$ of
 $\delta V^{(2)}$ reflect the $A$ dependence of the force strength
 $k^0$.  The interaction energy of $H^{\tau =0}_{\pi \nu}$ in the
 state with the total valence nucleon number $n$ and isospin $T$ is
 given by
\begin{eqnarray}
 U_{\pi\nu}^{\tau=0}= \langle H_{\pi\nu}^{\tau=0} \rangle =
 -\frac{1}{2}k^{0}\{ \frac{n}{2}(\frac{n}{2}+1)-T(T+1) \}.
 \label{eq:13}
\end{eqnarray}
\begin{small}
\begin{description}
\item{Table 1}{ The coefficient of empirical symmetry energy.}

\end{description}
\begin{center}
\begin{tabular}{|c|c|}\hline
 $a_{\rm sym}$ (MeV)  &      References                     \\ \hline
28.1    &   Ring and Schuck \cite{Ring}                          \\
23.3    &   Droplet model by Myers \cite{Myers}                  \\
25.0    &   Bohr and Mottelson \cite{Bohr}                       \\
  20.65 (1.0 - 2.32/$A^{1/3}$)  &   Green \cite{Green},  Seeger
   \cite{Seeger1}      \\
  30.59 (1.0 - 1.76/$A^{1/3}$)  &   Seeger \cite{Seeger2}        \\
  33.60 (1.0 - 1.52/$A^{1/3}$)  &   J.Duflo and A.P.Zucker
  \cite{Duflo} \\ \hline
\end {tabular}
\end{center}
\end{small}
If we calculate the double difference of the interaction energies
 $U_{\pi\nu}^{\tau=0}$ 
 (which we denote by $\delta^{(2)}U_{\pi\nu}^{\tau=0}$)
 by substituting $U_{\pi\nu}^{\tau=0}$ for $f$ in Eq. (\ref{eq:11}),
 we can expect 
 $\delta^{(2)}U_{\pi\nu}^{\tau=0} \approx \delta V^{(2)}$.
 We discussed in Ref. \cite{Kaneko3} that the average curve $40/A$ of
 $\delta V^{(2)}$ suggests the force strength $k^{0}= 80/A$ in the
 lowest order.  The simple form $k^{0}= 80/A$, however, cannot well
 reproduce the curve $40/A$ for $\delta^{(2)}U_{\pi\nu}^{\tau=0}$ as
 shown in Fig. 2.  The disagreement demands a higher order correction.
 According to our analysis \cite{Kaneko3}, the symmetry energy is also
 governed by the isoscalar {\it p-n} force $H^{\tau =0}_{\pi \nu}$.
 Let us look at the symmetry energy coefficient $a_{sym}=a(A)/4$
 in order to get a hint on the $A$ dependence of $k^0$.
 Table 1 lists the values of $a_{sym}$ proposed by various authors.
 The elaborate coefficient in Refs. \cite{Seeger1,Seeger2,Duflo,Green}
 has a mass dependence as the form $(1- \eta /A^{1/3})$
 (the lowest $A$ dependence $1/A$ of the symmetry energy is not
 included in the definition of $a_{sym}$).
 The following form of $k^0$, therefore, deserves to be tested:
\begin{eqnarray}
 k^{0} & = & \frac{\zeta}{A}(1.0 - \eta /A^{1/3}), \label{eq:14}
\end{eqnarray}
 where $\zeta$ and $\eta$ are parameters. 
 Figure 2 shows the values of $\delta^{(2)}U_{\pi\nu}^{\tau=0}$
 calculated with the use of the force strength (\ref{eq:14}).
 The calculated $\delta^{(2)}U_{\pi\nu}^{\tau=0}$ values finely
 reproduce the curve $40/A$ when we choose the parameters
 $\zeta=224.0$, $\eta=2.2$ for the $N,Z=20-50$ region and
 $\zeta=224.0$, $\eta=3.0$ for the $N,Z=50-82$ region.

 The four parameters $(g_{0},\chi,g_{2},k^{0})$ of our Hamiltonian 
 were fixed by Eqs. (7), (8) and (13). 
 We calculated the binding energies $E$ of $A=40-140$
 nuclei within the BCS approximation, i.e., the expectation values
 of the Hamiltonian with respect to the BCS states for the $A$ systems.
 (We did not calculate the binding energies near $N=Z$ nuclei
 and near the beginning and end of each shell region in order to avoid
 the error of the BCS approximation.)  The double differences of
 binding energies calculated from the approximate ground-state energies
 $E$ are shown in Figs. 3(a) and 3(b).  The calculated values are
 nicely around the curve $40/A$.  Especially, $\delta^{(2)}E$
 reproduces quite well the observed variation of $\delta^{(2)}V$ in
 Fig. 3(b).  We can say that our schematic Hamiltonian can
 qualitatively explain the behavior of the double differences of
 binding energies also in many-$j$ shell case.

 The double differences of binding energies, $\delta^{(1)}V$ and
 $\delta^{(2)}V$, are considered as good measures of the effective
 {\it p-n} interactions.  In the previous paper \cite{Kaneko3}, we
 investigated from what {\it p-n} inteactions the characteristic behavior
 of $\delta^{(1)}V$ and $\delta^{(2)}V$ come, in terms of the 
 single-$j$ shell model for $N \approx Z$ nuclei.  Let us develope the 
 analysis in the present many-$j$ shell calculation.
 We consider the contributions of respective forces to the ground-state
 energy in even-even, even-odd, odd-even and odd-odd nuclei,
 respectively,
\begin{eqnarray}
 E({e,e}) & = & \langle 0|H|0 \rangle =
  U_{0} + U_{\pi\nu}^{\tau=0}, \label{eq:15} \\
 E({e,o}) & = & \langle {n}|H|{n} \rangle =
  U_{0} + \Delta_{n} + U_{\pi\nu}^{\tau=0}, \label{eq:16} \\
 E({o,e}) & = & \langle {p}|H|{p} \rangle =
  U_{0} + \Delta_{p} + U_{\pi\nu}^{\tau=0}, \label{eq:17} \\
 E({o,o}) & = & \langle {pn}|H|{pn} \rangle =
  U_{0} + \Delta_{p} + \Delta_{n} + \epsilon_{pn}
   + U_{\pi\nu}^{\tau=0}. \label{eq:18}
\end{eqnarray}
 In the BCS approximation, the ground states of the four types of
 nuclei given above are defined by the quasiparticle vacuum, one 
 quasiparticle and {\it p-n} two quasiparticle states,
\begin{eqnarray}
|0 \rangle , \qquad
|{n} \rangle = a^{\dag}_{n}|0 \rangle , \qquad
|{p} \rangle = a^{\dag}_{p}|0 \rangle , \qquad
|{pn} \rangle =  a^{\dag}_{p}a^{\dag}_{n}|0 \rangle,
 \label{eq:21}
\end{eqnarray}
 We denote the ground-state energy of the even-even system $|0 \rangle$
 by $U_0$, the neutron (proton) gap by $\Delta_n$ ($\Delta_p$) and the
 interaction energy of the residual force ($QQ+P_2$) by
 $\epsilon_{pn}$,
\begin{eqnarray}
 U_{0} &=& \langle 0|H_{sp}+H_{P_{0}}+H_{QQ}+H_{P_{2}}|0 \rangle ,
  \label{eq:22} \\
 \Delta_\rho &=& \langle \rho | H_{sp}+H_{P_0}+H_{QQ}+H_{P_2}
  |\rho \rangle - U_0 \quad  (|\rho \rangle = |n \rangle \mbox{ or }
   |p \rangle ), \label{eq:221} \\
 \epsilon_{pn} & = & \langle {pn}|H_{QQ}+H_{P_{2}}|{pn}
  \rangle . \label{eq:23} 
\end{eqnarray}
 It is easy to calculate $\delta^{(1)}E$ and $\delta^{(2)}E$ from Eqs.
 (\ref{eq:15}-\ref{eq:18}).  We have seen in Fig. 2 that the
 value of $\delta^{(m)}U_{\pi\nu}^{\tau=0}$ yields the curve $40/A$,
 when we use the parameters ($\zeta$, $\eta$) fixed above.
 We therefore obtain the relation $\delta^{(1)}E({e,e})$=
 $\delta^{(1)}E({o,o})$=$40/A-\epsilon_{pn}$ for even $A$
 nuclei, $\delta^{(1)}E({e,o})$=$\delta^{(1)}E({o,e})$
 =$40/A+\epsilon_{pn}$ for odd $A$ nuclei and $\delta^{(2)}E=40/A$
 both for even and odd nuclei.  The characteristic behaviors of
 $\delta^{(1)}V$ and $\delta^{(2)}V$ are clearly understood.
 Their average values $40/A$ are attributed to the $J$ independent
 isoscalar {\it p-n} force also in the present many-$j$ shell calculation.
 The deviations from the average curve $40/A$ in $\delta^{(1)}V$
 are due to the residual p-n interaction between the last proton and
 neutron in odd-odd nuclei caused by the $QQ+P_2$ force (the $QQ$ 
 force mainly contributes to it).
 It is notable that the residual p-n interaction
 $\epsilon_{pn}$ is attractive.  This and the relation
 $E({o,o})-E({e,e})$=$\Delta_p+\Delta_n+\epsilon_{pn}$
 obtained from Eqs. (\ref{eq:15}) and (\ref{eq:18}) explain the fact
 that the binding energy difference $B({o,o})-B({e,e})$ is
 systematically smaller than the sum of gaps $\Delta_p+\Delta_n$
  (Bohr and Mottelson \cite{Bohr} suggested an attractive residual 
  {\it p-n}
 interaction about $20/A$ between the last odd proton and neutron
 in odd-odd nuclei).

  Let us now discuss the symmetry energy $E_{sym}$ and the Wigner
 energy $E_W$.  We calculated $E_{sym}$ and $E_W$ from the ground-state
 energies in the BCS approximation.  Figure 4 shows the calculated
 and experimental symmetry enrgy coefficient $a(A)$ in the expression
 $E_{sym}+E_W=a(A)T(T+1)/A$. (The $A$ dependence of $a(A)=4a_{sym}$
 extracted by Seeger \cite{Seeger2} and Duflo et al. \cite{Duflo} are
 also shown in Fig. 4.)
   The figure is plotted separately for
 "diagonal" and "off-diagonal" regions where neutrons and protons are
 in the same major shell for the former and in the different major
 shells for the latter.  We estimated the Coulomb-energy-corrected
 binding energies $B^{*}=B({\rm exp})+E_{\rm Coul}({\rm cal})$
 following Comay and J$\ddot{\rm a}$necke \cite{Comay} and extracted
 the symmetry energy coefficients by the treatment of
 J$\ddot{\rm a}$necke and Comay \cite{Janecke}.
 The agreement between theory and experiment is very well for the
 diagonal region of $N,Z=20-50$, and is qualitatively good for the
 off-diagonal region of $N,Z=20-50$ and the diagonal region of
 $N,Z=50-80$ where the calculated values are a little smaller than the
 experimental ones.
 The coefficient $a(A)$ in the off-diagonal region is approximately
 constant for the heavier nuclei as expected from the simple
 liquid-drop model. They are larger than those of the diagonal region,
 as pointed out by J$\ddot{\rm a}$necke and Comay \cite{Janecke}.
  In our calculation, the single-particle energy
 $E_{sp}=\langle 0|H_{sp}|0 \rangle$ influences the symmetry energy
 coefficient because of a large energy gap between the major shells.
 We can also see the shell-closure effect on the symmetry energy as
 well as the double difference of binding energies $\delta V^{(1)}$.
 
 We analyze the symmetry energy in the same way as the single-$j$
 shell model calculation \cite{Kaneko3}. 
 If the single-particle energies are roughly degenerate, i.e.,
 $\epsilon_a \approx \epsilon$, and $H_{QQ}+H_{P_{0}}$ can be
 neglected, the Hamiltonian $H_{sp}+H_{P_{0}}+H_{\pi \nu}^{\tau=0}$
 has the SO(5) symmetry.  Then, the ground-state energy is approximated
 in terms of the total valence nucleon number $n$ and the total isospin
 $T$ as follows:
\begin{eqnarray}
 \tilde{E} & = & \langle H_{sp} + H_{P_{0}} \rangle
  + U_{\pi\nu}^{\tau=0}  \nonumber \\
  & = & (\epsilon - \lambda)n
   - \frac{1}{2}\{ g_{0}n(\Omega - \frac{n-6}{4})
   + k^{0}\frac{n}{2}(\frac{n}{2}+1) \} \nonumber \\
  & {} & + \frac{1}{2}(g_{0}+k^{0})T(T+1).   \label{eq:24}
\end{eqnarray}
 The term proportional to $T(T+1)$ in the ground-state energy
 (\ref{eq:24}) just corresponds to $E_{sym}+E_W$.  The approximate
 symmetry energy coefficient $\tilde{a}(A)$ is given by
\begin{equation}
 \tilde{a}(A)= \frac{1}{2}(g_{0}+k^{0})A. \label{eq:25}
\end{equation}
 When we use the parameters $g_{0}=24/A$ and
 $k^{0}=224(1.0-2.2/A^{1/3})/A$ in the $fpg$ region, the symmetry
 energy coefficient in the above SO(5) limit becomes
 $\tilde{a}_{\rm sym}=\tilde{a}(A)/4=31(1.0-1.987/A^{1/3})$ from
 Eq. (\ref{eq:25}). This agrees considerably well with the empirical
 symmetry energy $30.59(1.0-1.76/A^{1/3})$ extracted by Seeger
 \cite{Seeger2}.  We can, therefore, say that the symmetry energy and
 Wigner energy are originated nearly in the $J$ independent isoscalar
 p-n force and the $J=0$ isovector pairing force.  The two
 contributions to $\tilde{a}_{\rm sym}$ are, for instance, 12.3 MeV
 and 3.0 MeV respectively in the $A=60$ system.  The isoscalar {\it p-n}
 force $H^{\tau =0}_{\pi \nu}$ plays a significant role in reproducing
 the experimental symmetry energy and the Wigner energy.
  
 In conclusion, the systematic behaviors of the double differences of
 binding energies, symmetry energy and Wigner energy are reproduced
 well by the schematic interaction $P_0+QQ+P_2+H^{\tau =0}_{\pi \nu}$
 within the BCS approximation, over the wide range of nuclei. 
 It was shown that the $J$ independent isoscalar p-n force is crucial
 for nuclear properties not only in $N \approx Z$ nuclei but also
 in $N>Z$ nuclei far from the $N=Z$ line.
 The analysis also tells that the residual interaction between the last
 proton and neutron in odd-odd nuclei consists in that of the $QQ$
 force. We estimated the effect of the ground-state correlation in 
 the framework of of the Random-Phase approximation. The calculations,
 however, did not change the conclusions mentioned above.

\newpage

\newpage
{\bf Figure captions}
\begin{description}
\item{Fig. 1}
 The double differences of binding energies derived from the
 experimental masses in the region $A=16-210$:
 (a) $\delta V^{(1)}(Z,N)$ as a function of $A=N+Z$;
 (b) $\delta V^{(2)}(Z,N)$ as a function of $A=N+Z$. The dots stand
 for even $A$ nuclei, and the crosses for odd $A$ nuclei.
 The curve $40/A$ is also plotted.
 
\item{Fig. 2} 
 The calculated values of $\delta^{(2)}U_{\pi\nu}^{\tau=0}(Z,N)$
  as a function of $A=N+Z$ in $N,Z$=20-50 and $N,Z$=50-82 regions.
  
\item{Fig. 3} 
 The calculated values of $\delta^{(1)}E(Z,N)$ and
 $\delta^{(2)}E(Z,N)$ as a function of $A=N+Z$ in $N,Z$=20-50
 and $N,Z$=50-82 regions. 

\item{Fig. 4} 
 The symmetry energy coefficient in $N,Z$=20-50 and $N,Z$=50-82
 regions. The calculated and experimental values are denoted by
 the open and filled circles, respectively. The full lines shows
 the empirical formula by Seeger \cite{Seeger2}, and the dotted line
 the empirical formula by Duflo and Zucker \cite{Duflo}.
\end{description}

\begin{figure}[h]
\begin{center}
   \includegraphics[height=20cm]{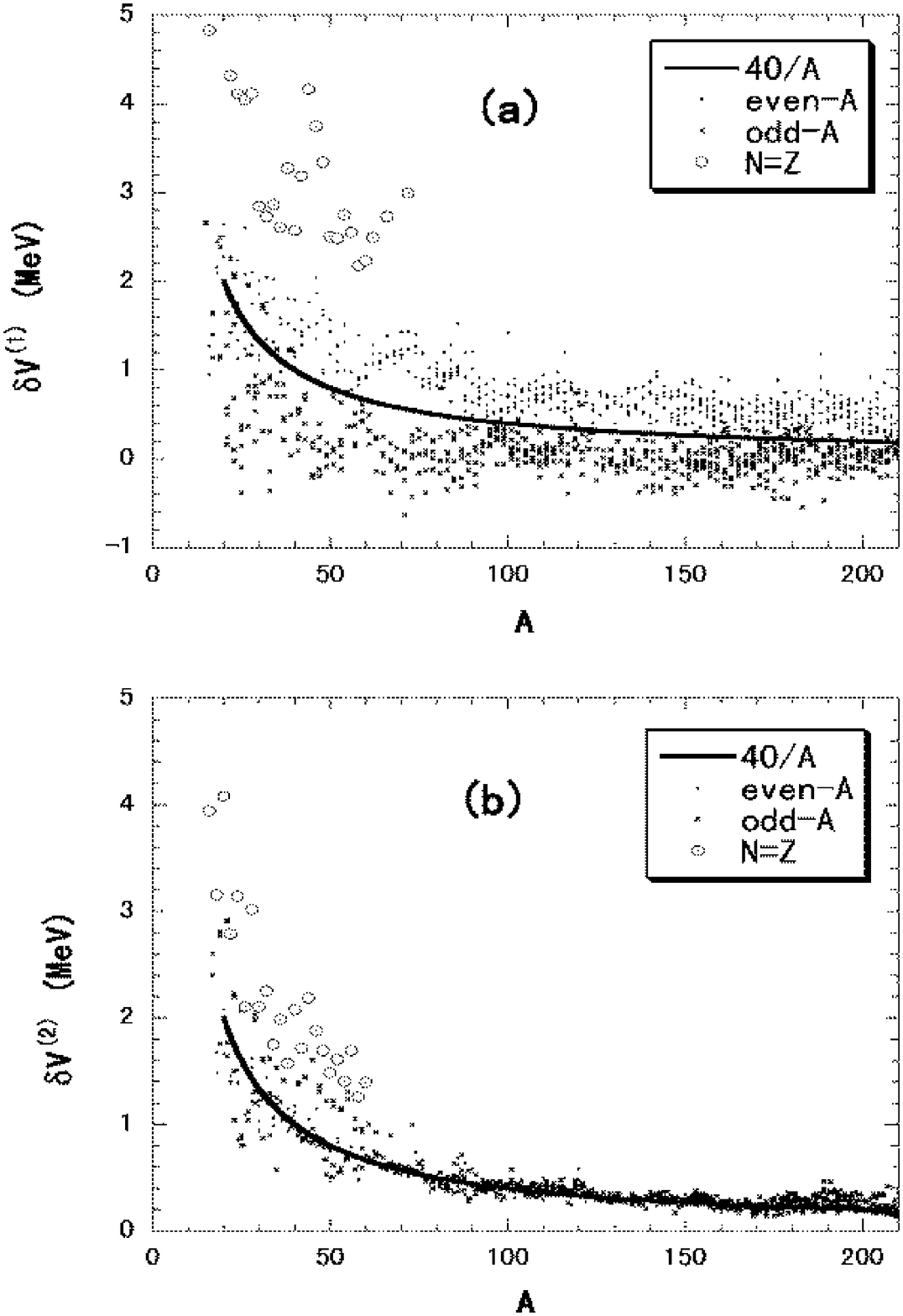}
\end{center}
\end{figure}
\begin{figure}[h]
\begin{center}
   \includegraphics[height=12cm]{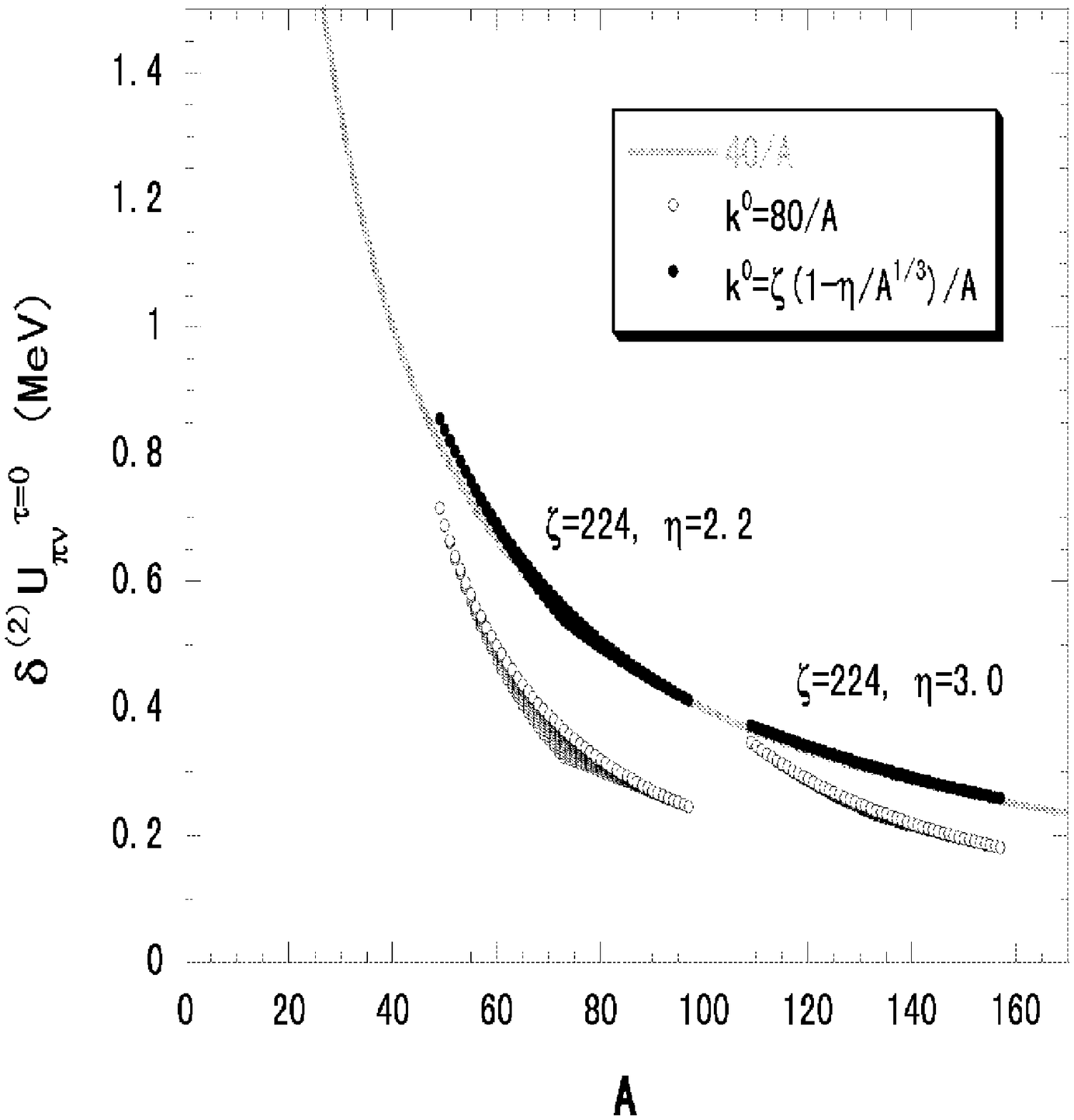}
\end{center}
\end{figure}
\begin{figure}[h]
\begin{center}
   \includegraphics[height=18cm]{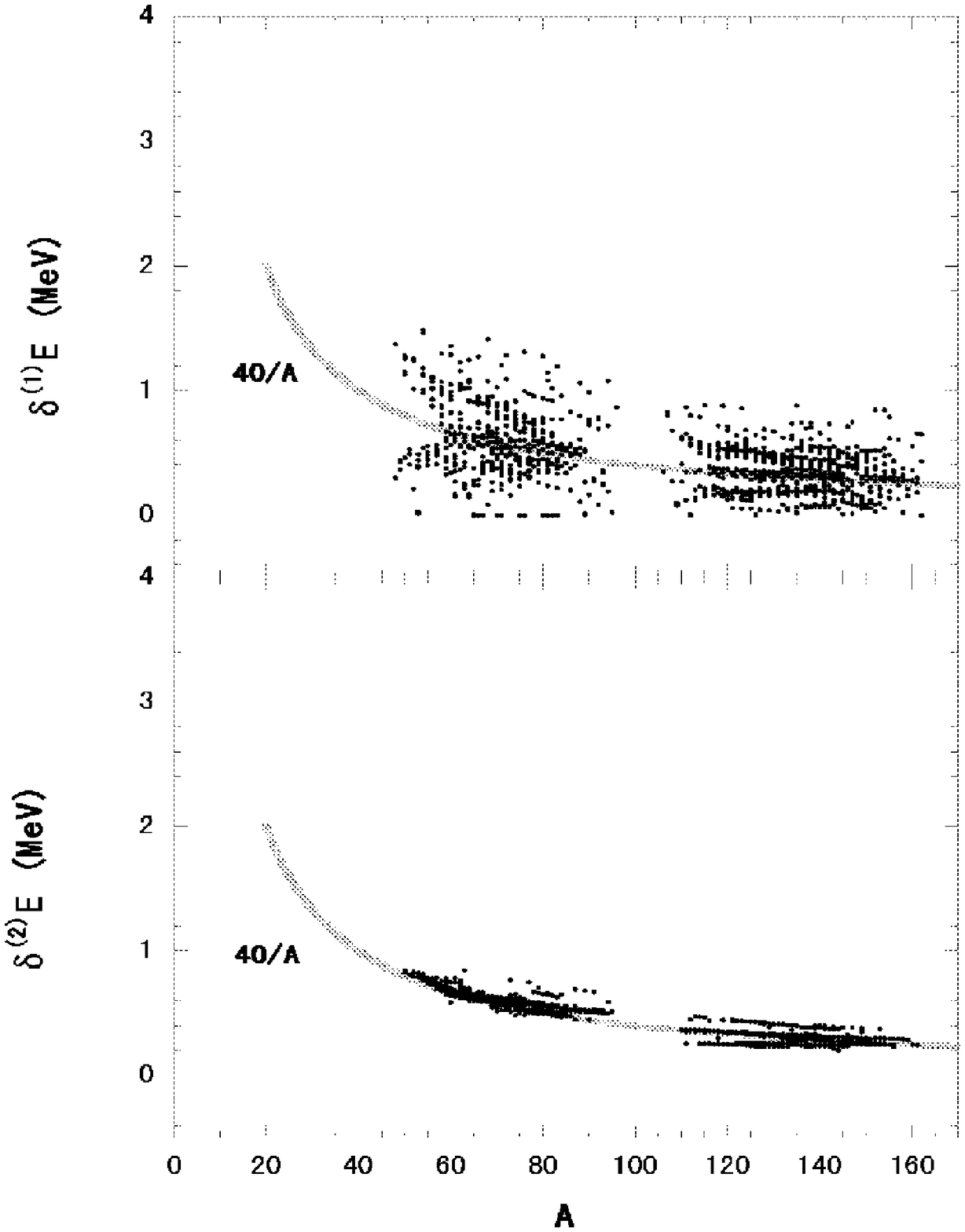}
\end{center}
\end{figure}
\begin{figure}[h]
\begin{center}
   \includegraphics[height=18cm]{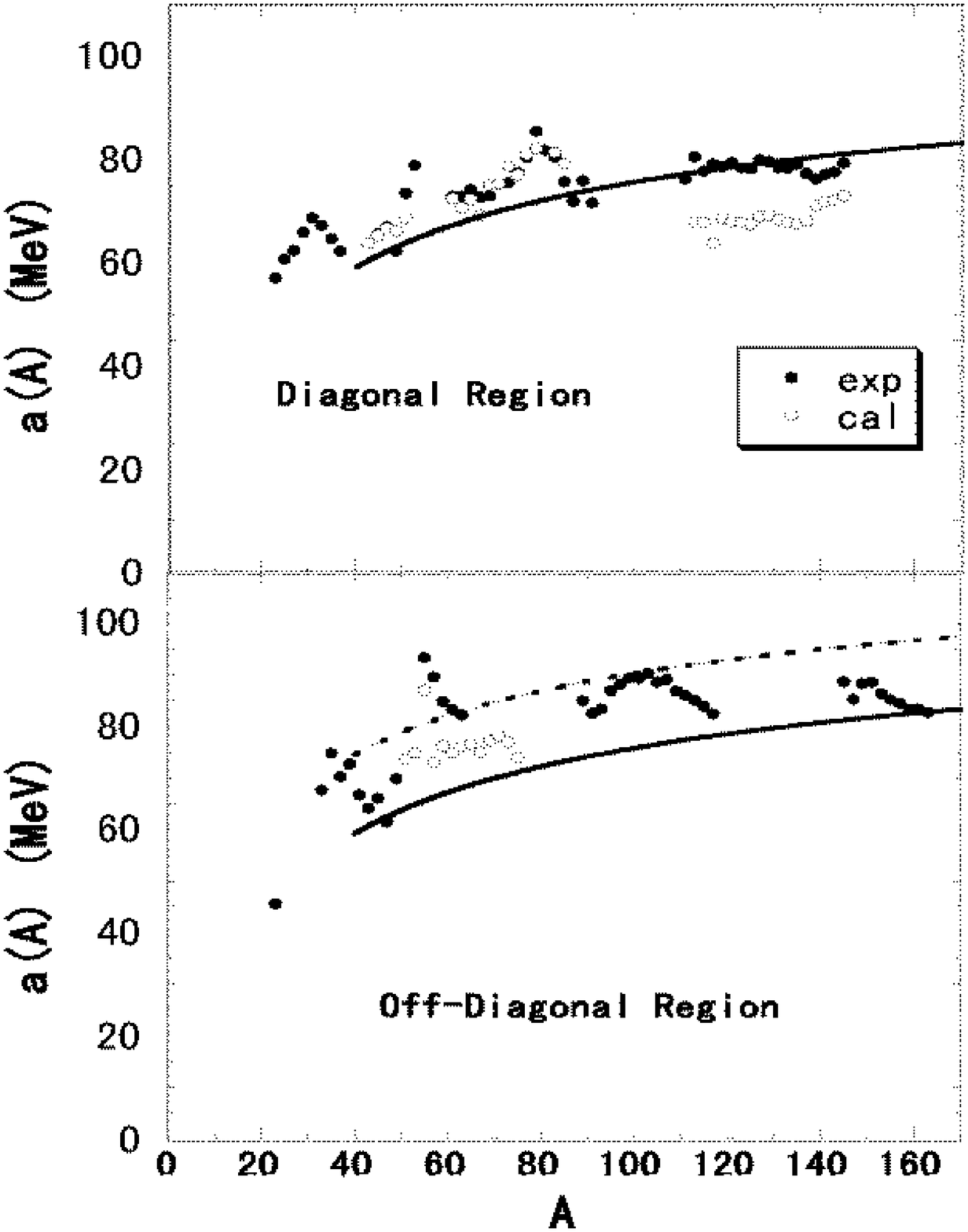}
\end{center}
\end{figure}

\makeatother

\end{document}